\documentclass{nature}
\usepackage{graphicx} 
\usepackage{mathtools}
\usepackage{amsfonts}
\usepackage{color}

\title{Single plasmon hot carrier generation in metallic nanoparticles}

\author{Lara Rom\'an Castellanos$^{1,3}$, Ortwin Hess$^{1,3}$ \& Johannes Lischner$^{2,3}$}

\begin{document}

\maketitle

\begin{affiliations}
 \item Department of Physics, Imperial College London, London SW7 2AZ
 \item Departments of Physics and Materials, Imperial College London, London SW7 2AZ
 \item The Thomas Young Centre for Theory and Simulation of Materials
\end{affiliations}

\begin{abstract}
Hot carriers produced from the decay of localized surface plasmons in metallic nanoparticles are intensely studied because of their optoelectronic, photovoltaic and photocatalytic applications. From a classical perspective, plasmons are coherent oscillations of the electrons in the nanoparticle, but their quantized nature comes to the fore in the novel field of quantum plasmonics. In this work, we introduce a quantum-mechanical material-specific approach for describing the decay of single quantized plasmons into hot electrons and holes. We find that hot carrier generation rates differ significantly from semiclassical predictions. We also investigate the decay of excitations without plasmonic character and show that their hot carrier rates are comparable to those from the decay of plasmonic excitations for small nanoparticles. Our study provides a rigorous and general foundation for further development of plasmonic hot carrier studies in the plasmonic regime required for the design of ultrasmall devices.
\end{abstract}

\section{Introduction}
Localized surface plasmons (LSPs) in metallic nanoparticles facilitate drastic electric field enhancements and large light absorption cross sections that can be harnessed in nanophotonic applications, such as plasmon-enhanced biosensing \cite{Li2015a}, surface-enhanced Raman scattering \cite{Willets2007}, data storage \cite{Ditlbacher2000} or nanoheaters \cite{Govorov2006,Huhn2011}.  Recently, there has been significant interest in the decay of the LSPs into electrons and holes. The resulting carriers are energetic or "hot" and can be used in solar energy conversion applications, including solar cells \cite{Hartland2017,Mubeen2013} or photocatalysts \cite{Clavero2014,Hartland2011,Brongersma2015}. For example, Mukherjee and coworkers demonstrated that hot electrons can induce challenging chemical reactions, such as the dissociation of hydrogen molecules on gold surfaces \cite{Mukherjee2013}. Moreover, the fast decay of LSPs can be used for new quantum information devices and in nanocircuitry \cite{Tame2013, Jacob2011}.

To provide insight and guidance in this rapidly evolving field, a detailed theoretical understanding of hot electron processes, including plasmon decay, hot carrier thermalization and recombination dynamics, is needed. Using semiclassical approaches, which combine a classical description of the LSP with a quantum-mechanical description of hot carriers, several groups analyzed the distribution of hot carriers resulting from the plasmon decay and studied its dependence on the nanoparticle size, material and environment \cite{Manjavacas2014,Forno2018,Brown2015,Sundararaman2014}. Providing general insight, the semiclassical approach is frequently based on a bulk dielectric function (such as a Drude model) and therefore cannot be used to describe small nanoparticles where quantum confinement effects play an important role \cite{Crut2017,Mittal2015}. In addition, it has been observed that non-plasmonic excitations, such as electron-hole pairs, can take place at similar energies as the plasmon resonance, but such excitations are not described accurately on the basis of a the semiclassical approach \cite{Zheng2015,Ma2015,Zhou2016}. Finally, a classical description of plasmons is less appropriate in the limit of low plasmon densities, where their quantized character must be captured \cite{Bergman2003,Lou2017,Celebrano2010}. Other groups have employed first-principles real-time time-dependent density-functional theory (TDDFT) to study plasmon \cite{Bonafe2017,Townsend2012} and hot carrier properties in small metallic nanoparticles \cite{Zhang2014,Kuisma2015,Ma2015,Rossi2017a}. While this method allows the study of nonlinear properties and scales favorably with the system size, it does not include a quantized treatment of the plasmon. Few attempts have been made to describe the effect of electron-plasmon interactions using quantized plasmons in metallic nanoparticles. Notably, Gerchikov and coworkers \cite{Gerchikov2002} and Weick et al. \cite{Weick2006} used a separation of centre-of-mass motion and relative motion of the electrons to derive a quantized electron-plasmon Hamiltonian. However, their approach cannot be used to study the decay of other neutral excitations, such as electron-hole pairs.  

In this paper, we thus present a fully quantum-mechanical approach to calculating the properties of hot carriers resulting from the decay of neutral excitations, such as LSPs or electron-hole pairs. In particular, we employ an effective Hamiltonian which describes the interaction of fermionic quasiparticles with bosonic neutral excitations and determine its parameters, including quasiparticle and plasmon energies and electron-plasmon coupling constants, using quantum-mechanical calculations. Most importantly, the electron-plasmon coupling strength is derived by comparing the electronic self energy of the effective Hamiltonian with the first-principles self energy within the GW approximation (where the electron self energy is approximated as the product of the electron Green's function G and the screened Coulomb interaction W). After identifying the dominant plasmonic and non-plasmonic neutral excitations, we first calculate the hot carrier generation rates for spherical nanoparticles with different radii and study the effect of quantum confinement on the hot carrier distributions. We find that a larger fraction of the plasmon energy is distributed to the electrons rather than the holes. Secondly, we compare hot carrier rates from the decay of plasmonic and non-plasmonic excitations. We also compare our quantum-mechanical results with semiclassical calculations and show that there is a significant discrepancy in the hot carriers rates for small nanoparticles.


\section{Results}

\subsection{Electron-plasmon coupling}

To describe the interaction between charged fermionic quasiparticles and neutral bosonic excitations, such as plasmons and electron-hole pairs, in metallic nanostructures, we employ the following effective Hamiltonian
\begin{equation}
\hat{H}_{\rm eff} = \sum_i \epsilon_i \hat{c}_{i}^{\dag}\hat{c}_{i} +  \sum_I \hbar \omega_I \hat{b}_{I}^{\dag}\hat{b}_{I} + \sum_{i,j,I} g^{I}_{ij} \, \hat{c}_{i}^{\dag}\hat{c}_{j} (\hat{b}_{I} + \hat{b}_{I}^{\dag}),
\label{eq: Heff}
\end{equation}
where $\hat{b}_I^{\dag}$ ($\hat{b}_I$) creates (annihilates) a neutral excitation in state $I$ with energy $\hbar \omega_I$ and $\hat{c}_i^\dag$ ($\hat{c}_i$) creates (destroys) a quasiparticle in state $i$ with energy $\epsilon_i$. Also, $g_{ij}^I$ is the electron-plasmon coupling, i.e. the matrix element that describes the scattering of quasiparticles from state $i$ into state $j$ via the emission or absorption of a collective excitation in state $I$. See Supplementary note 1 for the definition of the single plasmon operator within the second quantisation formalism.

A key challenge in using $H_{\rm eff}$ to describe metallic nanoparticles is the accurate determination of the various parameters, including energies of neutral and quasiparticle excitations and their coupling. Quasiparticle energies are formally defined as the poles of the one-electron Green's function and can be measured in photoemission and inverse photoemission experiments\cite{Martin2017}. Calculating the Green's function, for example via the GW method, is very challenging for metallic nanoparticles \cite{Varas2016} and therefore quasiparticle energies are often approximated using Kohn-Sham energies obtained from density-functional theory (DFT)\cite{Saito1990}. Similarly, energies of neutral excitations are defined as poles of a two-particle Green's function and can be obtained by solving the Bethe-Salpeter equation or from time-dependent density-functional theory (TDDFT). Plasmons arise because of the long-ranged nature of the Coulomb interaction. This is captured by the Hartree contribution to the total energy \cite{PinesDavid1963Eeis}, while exchange-correlation effects often play a minor role for plasmon properties. Neglecting exchange-correlation effects results in the well-known random-phase approximation (RPA) for neutral excitations\cite{Fetter1971}.  

\begin{figure}[htpb]
\centering
\includegraphics[width=9.5cm]{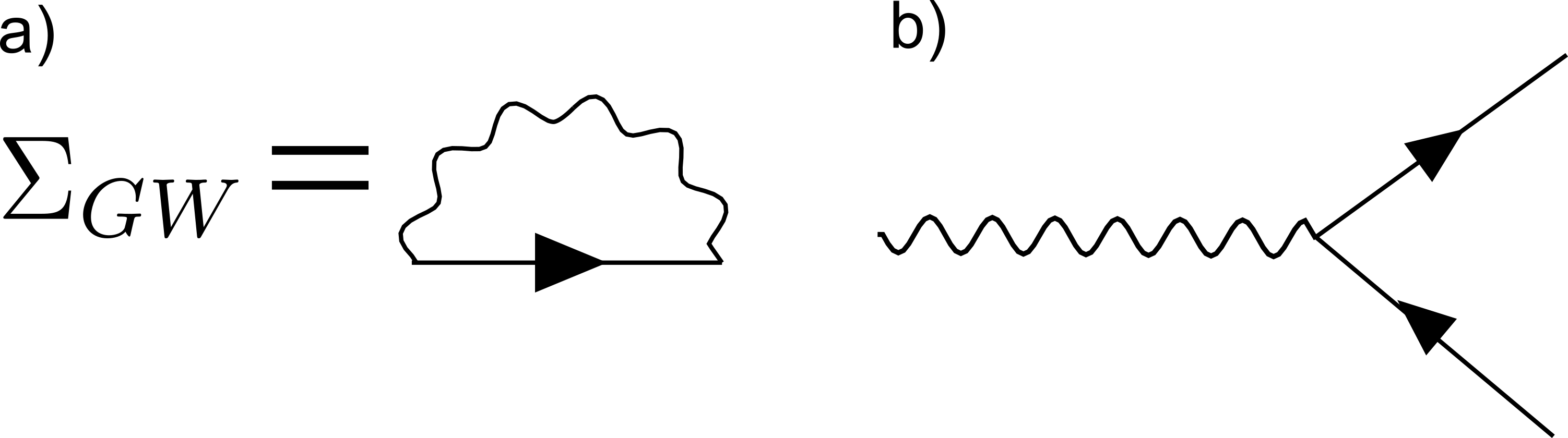}
\caption{Feynman diagrams. a) Ab initio GW self energy (b) Decay of a plasmon (or more generally, a neutral excitation) into an electron-hole pair. Solid lines indicate Green's functions $G$ that describe the propagation of fermionic quasiparticle excitations, such as quasi-electrons and quasi-holes, and wiggly lines denote the screened Coulomb interaction $W$, which describes the propagation of neutral bosonic excitations, such as plasmons or electron-hole pairs.}
\label{fig:plasmonFeynman}
\end{figure}

To determine the electron-plasmon coupling, we follow Lundqvist's approach \cite{Lundqvist1976} for the homogeneous electron gas and compare the second-order electron self-energy of $H_{\rm eff}$ with the correlation contribution of the ab initio GW self energy, see Figure \ref{fig:plasmonFeynman}a. The latter is given by \cite{Tiago2006}
\begin{equation}
\left< \phi_m \left| \Sigma^{\rm GW}(\omega) \right| \phi_m \right> = \sum_{j,I} \frac{|V^I_{mj}|^2}{\hbar \omega - \epsilon_j - \hbar \omega_I \eta_{j} - i\delta},
\label{eq:GWSigma}
\end{equation}
where the coefficient $\eta_j$ has the value $+1$ for unoccupied orbitals and $-1$ for occupied orbitals, $\delta$ represents a positive infinitesimal and $V^{I}_{mj}$ denotes the fluctuation potential given by 
\begin{equation}
 V^{I}_{mj} = \int d\mathbf{r} \int d\mathbf{r}' \phi_m(\mathbf{r})\phi_{j}(\mathbf{r})\frac{e^2}{|\mathbf{r} -\mathbf{r}' |}\rho_{I}(\mathbf{r}')
\label{eq:coupling}
\end{equation}
with the transition density 
\begin{equation}
\rho_{\rm I}(\mathbf{r}) = \sum_{vc} F^{I}_{\rm vc} \left(  \frac{\epsilon_{c}-\epsilon_{v}}{\hbar \omega_{I}} \right) ^{1/2}\phi_{\rm v}(\mathbf{r})\phi_{\rm c}(\mathbf{r})
\label{eq:CasidaAssignment}
\end{equation}
characterizing the $I$-th neutral excitation. Note that $\phi_v(\mathbf{r})$ ($\phi_c(\mathbf{r})$) denote single-particle wavefunctions of occupied (empty) states and the coefficients $F^I_{vc}$ are obtained by solving the Casida equation, see Methods section.

For the second-order electron self energy of $\hat{H}_{\rm eff}$, we find \cite{Mahan1981}
\begin{equation}
\left< \phi_m \left| \Sigma^{\rm eff}(\omega) \right| \phi_m \right> = \sum_{j,I} \frac{|g^I_{mj}|^2}{\hbar \omega - \epsilon_j - \hbar \omega_I \eta_{j}- i\delta}.
\label{eq:elecBosSigma}
\end{equation} 

Comparing Eq.~\eqref{eq:GWSigma} and Eq.~\eqref{eq:elecBosSigma} yields the final expression for the electron-plasmon coupling which is given by 
\begin{equation}
	g^I_{mj} = V^I_{mj}.
	\label{eq:g}
\end{equation}
This result has the intuitive interpretation that the coupling between the neutral excitation $I$ and the quasiparticles states $m$ and $j$ is due to the electric potential induced by the transition density of the neutral excitation, see Eq.~\eqref{eq:coupling}. Note that the expression is completely general and does not depend on the dimensionality, material or size of the system under consideration. 

\subsection{Hot carrier generation}
To model the decay of a neutral excitations, such as a plasmon, into hot electrons and holes, we calculate the self energy of the neutral excitation and evaluate the Feynman diagram shown in Figure~\ref{fig:plasmonFeynman}(b). The resulting decay rate $\Gamma_I$ is given by \cite{Forno2018}
\begin{equation}
\Gamma_{I} = \frac{2\pi}{\hbar} \sum_{vc} |g^I_{vc}|^{2} \delta(\epsilon_{c} - \epsilon_{v} - \hbar \omega_I).
\label{eq: rate}
\end{equation}
In our numerical calculations, the delta function was replaced by a Gaussian with a standard deviation of 0.12~eV. 

The generation rate $N^I_{e}(E)$ of hot electrons with energy $E$ from the decay of the $I$-th neutral excitation is given by 
\begin{equation}
N^I_e(E) = \frac{2\pi}{\hbar} \sum_{vc} |g^I_{vc}|^{2} \delta(\epsilon_{c} - \epsilon_{v} - \hbar \omega_I) \delta(E-\epsilon_c).
\end{equation}
A similar formula describes the generation rate of hot holes $N^I_h(E)$. In our numerical calculations, the second delta function was replaced by a Gaussian with a standard deviation of 0.05~eV.

Finally, the total generation rate $N^I_{\rm tot}$ of hot electrons resulting from the decay of the $I$-th collective excitation is given by\cite{Manjavacas2014}
\begin{equation}
N^I_{\rm tot} = \int_{E_F+ \delta E}^\infty dE N^I_e(E), \label{eq: Nhot}
\end{equation}
where $E_F$ denotes the Fermi energy which for metallic nanoparticles is defined as the middle of the gap between the highest occupied state and the lowest unoccupied one and $\delta E$ is a threshold energy which is typically chosen larger than the available thermal energy. As we are dealing with small nanoparticles, the energy gaps between occupied and unoccupied states are always larger than the thermal energy at room temperature and we therefore set $\delta E$ to zero. Note that the total rate of excited electrons is equal to the total rate of excited holes. 

\subsection{Semiclassical approach}

To model plasmon decay and hot carrier generation in nanoplasmonic systems, a semiclassical approach is usually employed \cite{Forno2018,Manjavacas2014}. In this approach, the metallic nanoparticle of radius $R$ is assumed to be exposed to an incident electric field (along the z-direction) with strength $E_0$. The total potential due to the perturbing field  and the induced polarization in the nanoparticle is calculated using the quasistatic approximation according to
\begin{equation}
\Phi_{sc}(\mathbf{r},\omega) = E_{0} \frac{\epsilon(\omega) - 1}{\epsilon(\omega) + 2} \left \{ 
\begin{tabular}{c} $r\cos \theta$ for $r \leq R$, \\
$R^{3} \frac{\cos \theta}{r^2}$ for $r > R$.
 \end{tabular}
 \right.
 \label{eq: SCpotential}
\end{equation} 
Here, $\epsilon(\omega)$ denotes the dielectric function of the bulk material which is often described using a Drude model
\begin{equation}
	\epsilon(\omega) = 1 - \frac{\omega_{0}^{2}}{\omega^2 + i\omega \gamma_P/\hbar},
	\label{eq: Drude}
\end{equation}
where $\gamma_P$ denotes the plasmon width and $\omega_{0}$ is the bulk plasmon frequency $\omega_{0} = \sqrt{4\pi n e^2/m}$ where $e$, $m$ and $n$ denote the electron charge, mass and density, respectively. Improved results can be obtained by using non-local approximations to the dielectric function \cite{Ciraci2012,Mortensen2014}.

The hot-carrier generation rates are then obtained by evaluating Fermi's Golden Rule in Eq.~\eqref{eq: rate} with the electron-plasmon coupling strength
\begin{equation}
g^{SC}_{ij} = \langle \phi_i | \Phi_{sc} | \phi_j \rangle.
\label{eq: SCcoupling}
\end{equation}

Importantly, the semiclassical expression for the electron-plasmon coupling depends on the strength of the incident light field which excites the plasmon. In contrast, the quantum definition, Eq.~\eqref{eq:g}, does not depend on $E_{\rm 0}$. To resolve this discrepancy, we note that the semiclassical result contains the interaction of the electrons with both the induced charge density and the external light field, while the quantum result only describes the interaction with the (light induced) neutral excitations. Moreover, the semiclassical result captures the effect of exciting multiple plasmons (in fact, the expression becomes exact in the limit of a large number of plasmon quanta), while the quantum result describes the decay of a single plasmon excitation.

In order to meaningfully compare the results of the two approaches, we derive an expression for the electric field strength $E^{1PL}_0$ which is required to excite a single plasmon, see Supplementary note 2. We find that
\begin{equation}
	E^{1PL}_0 = \frac{\gamma_P}{\mu_P},
\label{eq: E_0}	
\end{equation}
where $\mu_P$ denotes the dipole moment of the plasmonic state. The corresponding semiclassical transition dipole moment is given by $\mu_{SC}(\omega) =   R^{3} \left(\frac{\epsilon(\omega) - 1}{\epsilon(\omega) + 2} \right) E^{1PL}_{0} $. When evaluated at the classical Mie frequency $\omega_{cl}=\omega_0/\sqrt{3}$, we obtain $\mu_{SC}(\omega_{cl}) = R^{3}\hbar \omega_{cl} / \mu_{P}$, which is independent of the plasmon linewidth $\gamma_P$.

\section{Numerical results}

We present results for three sodium nanoparticles: Na$_{40}$, Na$_{58}$ and Na$_{92}$, which consist of 40, 58 and 92 Na atoms, respectively. These systems are modelled as jellium spheres with diameters of 1.4~nm, 1.6~nm and 1.9~nm, respectively. We first analyze the neutral excitations of these systems and identify those with plasmonic character and then calculate the hot carrier distributions resulting from their decay. We also compare our results to semiclassical calculations and obtain hot carrier rates for nanoparticles in different dielectric environments.

\subsection{Identifying plasmonic excitations}
Distinguishing plasmon-like excitations, which are a typically thought of as collective oscillations of all electrons in the nanoparticle, from other neutral excitations, such as bound or unbound electron-hole pairs, is difficult. Recently, several approaches have been proposed to address this problem, for example by scaling the electron-electron interactions\cite{Casanova2016,Bernadotte2013}, by studying the time-dependent occupations of the Kohn-Sham states\cite{Townsend2014} or by investigating the optical response of the nanostructure\cite{Zhang2017b,Bursi2016} that led to the introduction of the generalized plasmonicity index. Below, we present an alternative method for identifying the plasmon which is based on a graphical analysis of various physically transparent quantities (such as the excitation energy and oscillator strength of a neutral excitations) which are directly obtained from a solution of Casida's equation. Importantly, this approach does not require multiple solutions of Casida's equation at different strengths of the electron-electron interaction and shares many advantages of the generalized plasmonicity index, which is proportional to the total induced dipole moment of a transition.

Importantly, our quantum method for calculating hot-carrier rates can be used for both plasmon-like and electron-hole pair-like excitations. However, to enable a comparison with semiclassical calculations, we have employed the following approach to identify excitations with a plasmonic character. In particular, we analyze three properties of neutral excitations: (i) the energy $\hbar \omega_I$ of the excitation, (ii) its oscillator strength $f_I$ and (iii) its collectivity $C_I$, see Methods section. For plasmon-like excitations, we expect both high oscillator strengths and collectivities as well as energies not too far from the classical plasmon energy $\hbar \omega_{cl}$. For sodium nanoparticles, the classical plasmon energy is given by $\hbar \omega_{cl}=3.40$~eV.  

\begin{figure}[htpb]
\centering
\includegraphics[width=200pt]{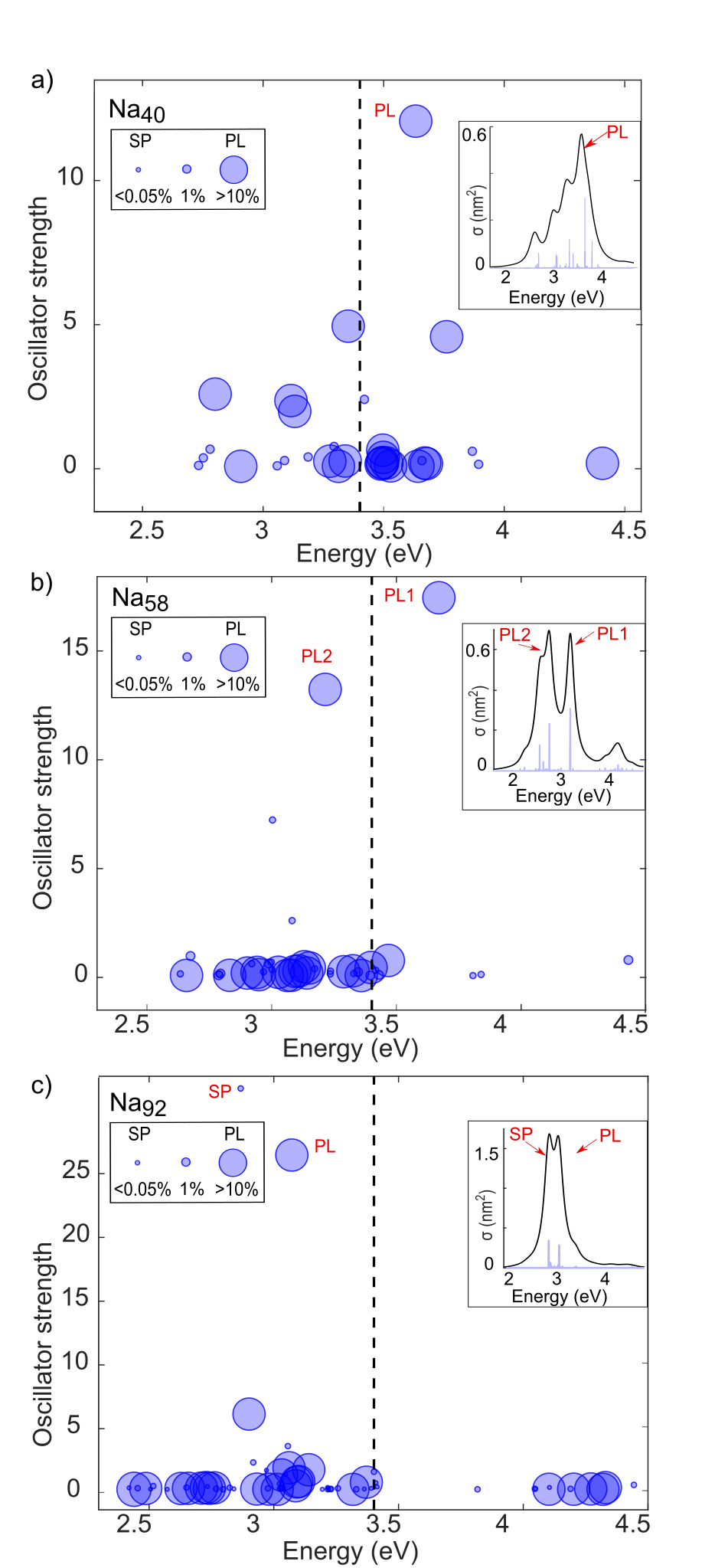}
\caption{Characterization of neutral excitations. (a) Neutral excitations in Na$_{40}$. (b) Neutral excitations in Na$_{58}$. (c) Neutral excitations in Na$_{92}$. Each neutral excitation is represented by a circle. The radius of each circle is proportional to the collectivity of the excitation. Plasmonic excitations (denoted PL) are expected to have a high oscillator strength and collectivity as well as energies close to the classical Mie plasmon energy (denoted by the dashed vertical line). In contrast, electron-hole pair excitations (denoted SP) have a low collectivity. The insets show the corresponding optical absorption spectra.}
\label{fig: Bubbles}
\end{figure}

\begin{figure}[htpb]
\centering
\includegraphics[width=472pt]{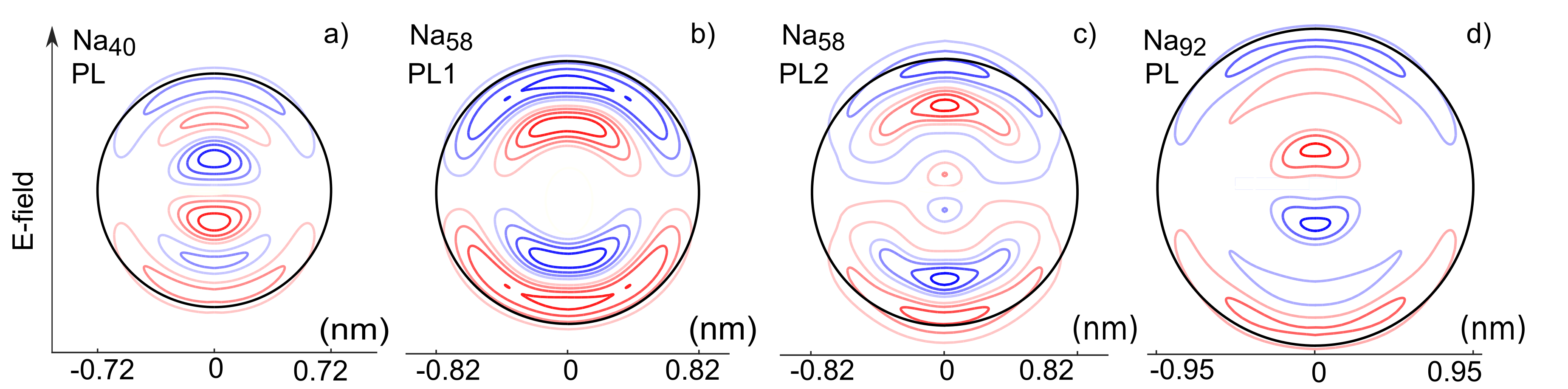}
\caption{Transition densities of plasmonic excitations (denoted PL). (a) Plasmonic charge density for PL state in Na$_{40}$. (b) Plasmonic charge density for PL1 state in Na$_{58}$. (c) Plasmonic charge density for PL2 state in Na$_{58}$. (d) Plasmonic charge density for PL state in Na$_{92}$. Note that Na$_{58}$ exhibits two neutral excitations with a plasmon-like behaviour.}
\label{fig: TransDensSize}
\end{figure}

Figure~\ref{fig: Bubbles} a)-c) summarizes the properties of neutral excitations for the three sodium nanoparticles under consideration. In these graphs, each circle represents an excited state with its radius being proportional to $C_{I}$. For Na$_{40}$, we find one state (denoted PL for plasmon in Figure\ref{fig: Bubbles}a) with a high oscillator strength, large collectivity and energy close to $\hbar \omega_{cl}$. This state gives rise to a strong peak in the absorption spectrum (see inset) and its transition density has a dipolar shape, see Figure~\ref{fig: TransDensSize}. We therefore assign this state a plasmonic character. For Na$_{58}$, we find two states with large oscillator strengths and high collectivities. The energies of both states are within 0.3~eV from $\hbar \omega_{cl}$ and their transition densities are both plasmon-like. We therefore assign both excitations plasmonic character (and denote them by PL1 and PL2 in in Figure\ref{fig: Bubbles}b)).  For Na$_{92}$, we find two states with high oscillator strengths. The state with energy $\hbar \omega_{SP} = 2.87$~eV has a low collectivity and therefore electron-hole pair character (denoted SP for single pair in in Figure \ref{fig: Bubbles}c)), while the state with energy $\hbar \omega_P=3.07$~eV has a high collectivity and plasmonic character (denoted PL in the figure). 

Table~\ref{tbl: Nhot} shows an overall redshift of the plasmonic excitation energy as the nanoparticle size increases. This is caused by quantum confinement effects and the spill-out of the electron wave functions beyond the geometrical radius of the nanoparticle \cite{Weick2006}. However, the redshift is not monotonic which has also been observed experimentally for nanoparticles with a diameter smaller than 3.5~nm \cite{Scholl2012, Reiners1995} and was predicted theoretically for small nanoparticles \cite{Weick2006}. Moreover, the presence of more than one plasmonic resonance for a given nanoparticle could explain the experimentally observed scattered distribution of energies in this size regime \cite{Scholl2012}. 

\begin{figure}[htpb]
\centering
\includegraphics[width=236.84 pt]{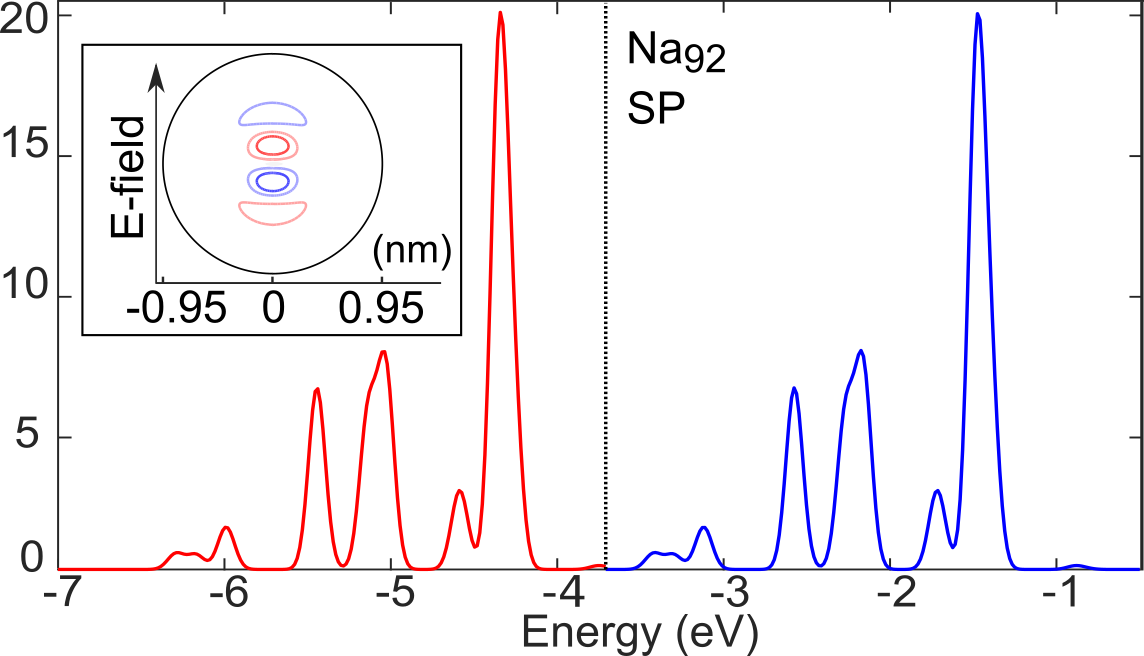}
\caption{Hot carrier distribution for a single-pair state.
Energy distribution of hot electrons and holes from the decay of the dominant electron-hole pair excitation in Na$_{92}$. The dashed line denotes the Fermi energy. The inset shows the transition density of the electron-hole pair excitation.}
\label{fig: HCD92SP}
\end{figure}

Figure~\ref{fig: TransDensSize} shows the transition densities of the plasmonic excitations in Na$_{40}$, Na$_{58}$ and Na$_{92}$. The transition densities exhibit a clear dipolar shape, but are significantly more complex than predicted by classical electrodynamics. In particular, the transition densities are not only localized at the surface of the nanoparticles and show multiple regions of positive and negative charge. While the transition density of the smallest system (Na$_{40}$) is extended throughout the nanoparticle, it becomes more localized in the surface regions for the larger nanoparticles in agreement with the classical result. In contrast, the transition densities of single-pair excitations can exhibit different features. For example, the transition density of the dominant single-pair state of Na$_{92}$, see inset of Figure~\ref{fig: HCD92SP}, is localized in the centre of the nanoparticle.

\subsection{Hot carrier distributions}

\begin{figure}[htpb]
\centering
\includegraphics[width=200.84pt]{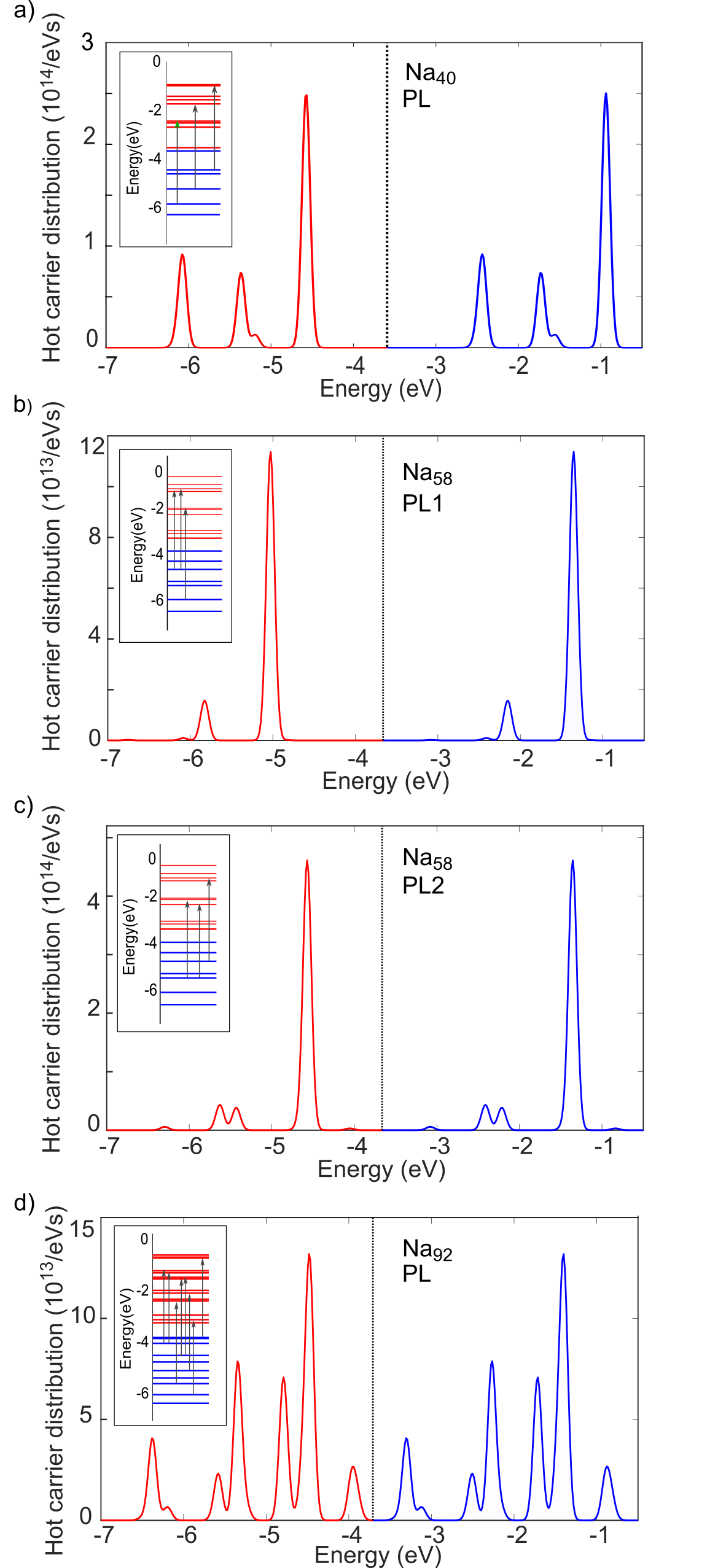}
\caption{Energy distributions of hot electrons and holes from the decay of plasmonic excitations. (a) From the decay of the plasmonic excitation in Na$_{40}$. (b) From the decay of the plasmonic excitation PL1 in Na$_{58}$. (c) From the decay of the plasmonic excitation PL2 in Na$_{58}$. (d) From the decay of the plasmonic excitation in Na$_{92}$. Blue lines for electrons and red lines for holes. The dashed lines denote the Fermi energy. The insets show the corresponding electronic transitions from occupied (blue) to empty (red) levels.}
\label{fig: HCDALL}
\end{figure}

Figure~\ref{fig: HCDALL} shows the energy distribution of electrons and holes resulting from the decay of a plasmon in the three sodium nanoparticles under consideration. Energy conservation imposes that the distribution of electrons (blue curve) has the same shape as the distribution of holes (red curve), but shifted by the plasmon energy. For the smaller nanoparticles, Na$_{40}$ and Na$_{58}$, the distributions exhibit only a few peaks. This is a consequence of the large energy level spacing (due to quantum confinement) which makes it difficult to find energy-conserving transitions, see insets of Figures.~\ref{fig: HCDALL}~a)-d). In contrast, a large number of peaks are observed in the hot carrier distributions of Na$_{92}$, where the energy level spacing is significantly reduced. Interestingly, the larger fraction of the plasmon energy is transferred to the hot electrons in these systems. 

\begin{table}[h]
  \caption{Excitation energies, total hot-carrier generation rates and transition dipole moments $\mu_I$ of the dominant neutral excitations in small sodium nanoparticles. SP refers to electron-hole pair excitations and PL to plasmonic excitations.}
  \begin{tabular}{ccccc}
    \hline
    System  & Excitation  & $\hbar \omega_{I}(eV)$ & $N_{\rm tot}$ ($1/ps$) & $\mu_I$ ($10^{-20}$~C nm)\\
    \hline\hline
    Na$_{40}$ & PL & 3.63 & 27.2 & 5.6  \\
    \hline
    Na$_{58}$ & PL1 & 3.67 & 8.2 & 6.8 \\
    Na$_{58}$ & PL2 & 3.21 & 35.2 & 6.3 \\
    \hline
    Na$_{92}$ & PL & 3.07 & 25.7 & 9.1 \\
    Na$_{92}$ & SP & 2.87 & 30.3 & 10.3 \\
    Na$_{92}$ & Classical PL & 3.40 & 976.8 & 57.7 \\
    \hline
  \end{tabular}
    \label{tbl: Nhot}
\end{table}

Table~\ref{tbl: Nhot} shows the total number of hot carriers generated by the decay of a single excitation in the three Na nanoparticles. The total generation rate for the dominant plasmonic excitation (PL in Na$_{40}$ and Na$_{92}$ and PL2 in Na$_{58}$) does not show a strong dependence on the nanoparticle radius. It is important to recall, however, that these rates are calculated for a single excitation quantum. As the transition dipole moment of these excitations increases with system size, see Table~\ref{tbl: Nhot}, more excitation quanta are excited per incoming photon and this gives rise to a larger number of hot carriers in larger nanoparticles.

As discussed in the Methods section, our approach can also be used to study the hot-carrier distribution resulting from the decay of neutral excitations without plasmonic character. Figure~\ref{fig: HCD92SP} shows the hot-carrier distributions resulting from decay of the prominent single-pair excitations in Na$_{92}$ (denoted SP in Figure~\ref{fig: Bubbles}~c). Similar to the plasmonic excitation in this system, the larger fraction of the excitation energy is transferred to the hot electrons. Interestingly, the total rate of hot carriers from the decay of the electron-hole pair excitation is larger than from the decay of the plasmonic excitation in this system, see Table~\ref{tbl: Nhot}.

\begin{figure}[htpb]
\centering
\includegraphics[width=472pt]{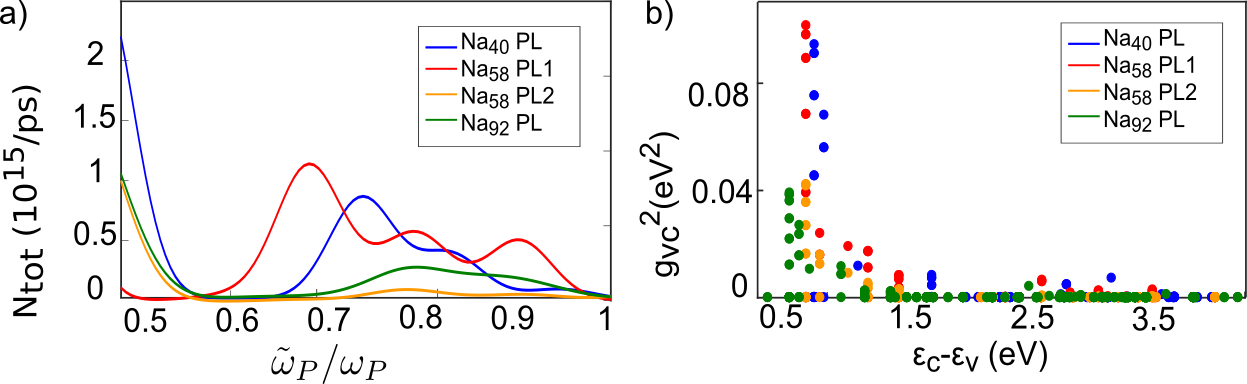}
\caption{Plasmonic hot carriers in different dielectric environments. (a) Total number of plasmonic hot carriers in different dielectric environments. Here, $\tilde{\omega}_P$ denotes the plasmon energy of the small sodium nanoparticle in the dielectric environment and $\omega_P$ denotes the corresponding value in vacuum. (b) Squared electron-plasmon coupling strengths versus the energy of the transition.}
\label{fig: NtotVswp}
\end{figure}

The energy of the localized surface plasmon can be easily modified by placing the nanoparticle in different dielectric environments. Specifically, the plasmon energy is reduced as the dielectric constant of the environment increases. To study the effect of the dielectric environment on the hot-carrier generation rates resulting from the decay of plasmonic excitations, we evaluated Eq.~\eqref{eq: rate} with a reduced plasmon energy $\tilde{\omega}_P$ (but keeping the orbital energies and coupling strengths unchanged). Figure~\ref{fig: NtotVswp} shows the resulting total hot-carrier generation rates for three Na nanoparticles as function of the environment-screened plasmon energy. We observe that the reduction of the plasmon energy can lead to significant enhancements in the total hot-carrier rates when the reduced plasmon energy matches the energy of multiple transitions from occupied to empty states in the quasiparticle spectrum. A particularly strong increase is found for plasmon energies smaller than half of the unscreened value $\omega_P$. This is caused by a strong increase of the electron-plasmon coupling strength for low-energy transitions, see inset of Figure~\ref{fig: NtotVswp}.

\subsection{Comparison to the semiclassical approach}

To compare the hot-carrier rates from our quantum approach to the semiclassical approximation, we carry out semiclassical calculations using the electric field strength $E_{\rm 0}^{1PL}$ which generates a single plasmon in the nanoparticle. Table~\ref{tbl: Nhot}
shows the resulting total hot-carrier rate for Na$_{92}$ for the classical plasmon energy. Surprisingly, we find that the semiclassical rate is more than an order of magnitude larger than the quantum result. 

To understand this discrepancy, we compare the transition dipole moment of the LSP obtained from the quantum calculation with the results from the semiclassical approach. Table~\ref{tbl: Nhot} shows that the semiclassical transition dipole moment of the plasmon is six times larger than the quantum mechanical one. This is consequence of two factors: (i) as discussed above, the transition density of the plasmons in small nanoparticles deviates significantly from the perfect dipolar shape predicted by classical electrodynamics (see Figure~\ref{fig: TransDensSize}) and (ii) in small nanoparticles, the plasmon is one of many excitations which share the total available oscillator strength (according to the f-sum rule), while in the semiclassical approach the total oscillator strength is concentrated in the plasmon, see Supplementary note 3. Because of their smaller transition dipole moments, the quantum plasmons couple less strongly to the transition dipole moments of the hot electron-hole pairs resulting in significantly reduced generation rates compared to the semiclassical calculations.

\section{Discussion}

We have presented a new approach for studying hot carrier generation in metallic nanoparticles which takes quantum plasmonic effects, such as the bosonic nature of the plasmon, fully into account. We employ an effective fermion-boson Hamiltonian and determine its parameters for specific nanoparticles using (time-dependent) density-functional theory and many-body perturbation theory within the GW approximation. In particular, an expression for the coupling strength of quasiparticles to neutral excitations is obtained by comparing the self energy of the effective Hamiltonian with the first-principles GW self energy. 
We have used this approach to study the decay of single plasmons into hot carriers in small sodium nanoparticles with different radii. We find that some systems exhibit multiple plasmonic excitations, while others have electron-hole pair excitations with large oscillator strengths. The hot carrier distributions from the decay of these excitations exhibit a molecular character with discrete peaks that are more closely spaced as the nanoparticle size increases. Interestingly, we find that in all systems a large fraction of the plasmon energy is transferred to the hot electrons. We also compare our results to semiclassical calculations and find that the semiclassical results provide qualitative insights but overestimate hot carrier rates for the small nanoparticles under consideration. Our approach opens the possibility to study hot carrier generation in the quantum plasmonic regime with potential application to quantum-controlled devices, including single-photon sources, transistors and ultra-compact circuitry at the nanoscale.

\subsection{Methods}
In this section, we describe how the various parameters of the effective electron-plasmon Hamiltonian $H_{\rm eff}$, Eq.~\eqref{eq: Heff}, are obtained. 

\subsection{Quasiparticle energies}

To model the electronic structure of spherical metallic nanoparticles of radius $R = r_{\rm s} N^{1/3}$ (where $r_{\rm s}$ denotes the Wigner-Seitz radius of the bulk material and $N$ is the number of electrons in the nanoparticle), we employ the jellium approach which has been widely used to describe metallic clusters \cite{Ekardt1984a,Brack1993,Beck1984,Bertsch1990}. In this parameter-free method, the positive charge of the atomic nuclei is smeared out homogeneously throughout the volume of the nanoparticle.The jellium approach often yields accurate electronic properties for metals with s- or p-electron bands, but at a significantly reduced computational cost compared to full atomistic descriptions. Other authors \cite{Manjavacas2014,Martins1981} have used phenomenological spherical well models to describe the electronic structure of metallic nanoparticles, but we have found that such models do not reproduce the experimentally observed ordering of energy levels \cite{Wrigge2002}when electron-electron interactions are included. For example, photoelectron spectroscopy of sodium clusters reveals the following ordering of energy levels (in order of increasing energy): 1s, 1p, 1d, 2s, 1f and 2p \cite{Wrigge2002}. This ordering is reproduced by the jellium approach, but not by the interacting spherical well approach.

Ground state properties, such as the total energy or the electron density, are obtained by solving the Kohn-Sham equations
\begin{equation}
\left[ -\frac{\hbar^2}{2m}\nabla^{2} + V_{KS}(\mathbf{r}) \right]  \phi_{i}(\mathbf{r}) = \epsilon_{\rm i} \phi_{i}(\mathbf{r}),
\label{eq:Schr_complete}
\end{equation}
where $\phi_i(\mathbf{r})$, $V_{KS}(\mathbf{r})$ and $\epsilon_i$ denote the Kohn-Sham orbitals, Kohn-Sham potential and Kohn-Sham energies, respectively. The Kohn-Sham potential consists of a nuclear, a Hartree and an exchange-correlation contribution, for which we employ the Perdew-Zunger parametrization of the local density approximation (LDA) \cite{Perdew1981}.

\begin{figure}[htpb]
\centering
\includegraphics[width=236pt]{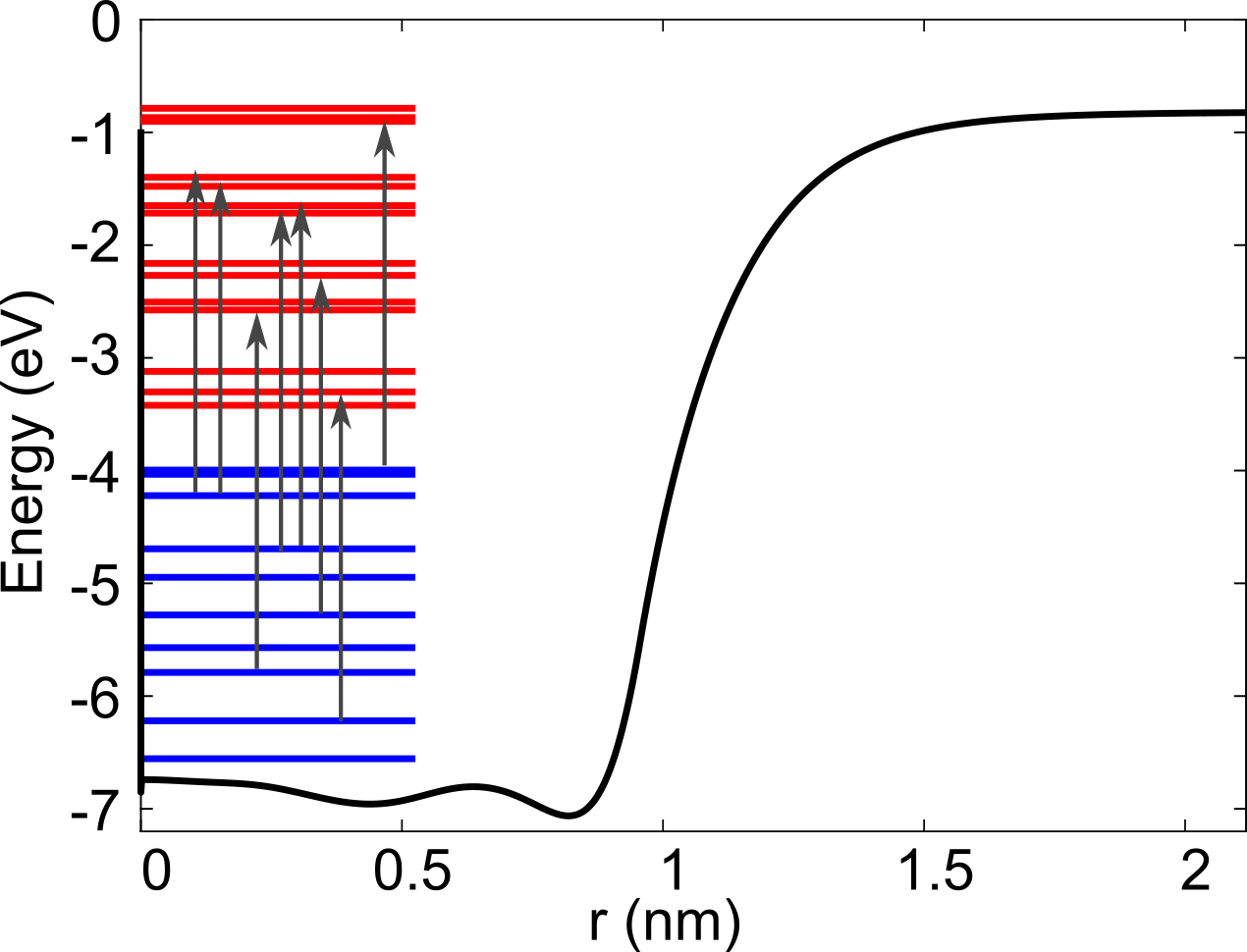}
\caption{Ground state of the nanoparticle. Kohn-Sham potential (shifted by $\Delta$-SCF correction) and quasiparticle energy levels of a sodium nanoparticle consisting of 92 atoms. Occupied orbitals are denoted by blue lines and unoccupied orbitals by red lines. Calculations are carried out using the jellium approach.}
\label{fig: Jellium}
\end{figure}

We only carry out calculations for nanoparticles with closed electronic shells. For such systems, the Kohn-Sham potential exhibits spherical symmetry and the Kohn-Sham orbitals can be expressed as the product of a spherical harmonic and a radial function which is obtained by direct integration on a real-space grid. To approximate quasiparticle energies which correspond to electron addition or removal energies, we have calculated the ionization potential of the nanoparticles using the $\Delta$-SCF approach (i.e., by calculating the total energy difference of the neutral and ionized nanoparticles) and shifted all Kohn-Sham energies such that the energy of the highest occupied orbital agrees with the calculated ionization potential. Alternatively, the quasiparticle energies can be obtained from GW calculations, but at a significantly larger computational cost. Figure~\ref{fig: Jellium} shows the resulting quasiparticle energy levels and Kohn-Sham potential for a sodium nanoparticle ($r_s=4.00$~Bohr) consisting of 92 atoms.

\subsection{Neutral excitations}

Neutral excitations are obtained using the random phase approximation (RPA). In particular, we solve Casida's equation \cite{Casida1995} 
\begin{equation}
\sum_{v'c'} \Omega_{vc,v'c'} F^{\rm I}_{v'c'} = \hbar^2 \omega_{\rm I}^{2} F^{\rm I}_{vc},
\end{equation}
where $\hbar\omega_I$ denotes the energy of the neutral excitations and $F^I_{vc}$ is the corresponding eigenvector, which determines the transition density, see Eq.~\eqref{eq:CasidaAssignment}. The Casida matrix is given by 
\begin{equation}
\Omega_{vc, v^{\prime} c^{\prime}} = \delta_{\rm cc^{\prime} } \delta_{\rm v v^{\prime}} (\epsilon_{\rm c} - \epsilon_{\rm v})^{2} + 2 \sqrt{f_{vc}(\epsilon_{\rm c} - \epsilon_{\rm v})} K_{vc, v^{\prime} c^{\prime}} \sqrt{f_{v^{\prime}c^{\prime}}(\epsilon_{\rm c^{\prime}} - \epsilon_{\rm v^{\prime}})}, 
\label{eq:CasidaOmega}
\end{equation}
where $K_{vc, v^{\prime} c^{\prime}}$ denote Coulomb matrix elements 
\begin{equation}
K_{vc, v^{\prime} c^{\prime}} =  \int d\mathbf{r} \int d\mathbf{r}^{\prime} \phi_{\rm v}(\mathbf{r}) \phi_{\rm c}(\mathbf{r}) \frac{e^2}{\vert \mathbf{r} - \mathbf{r}^{\prime} \vert} \phi_{\rm v^{\prime}}(\mathbf{r}^{\prime}) \phi_{\rm c^{\prime}}(\mathbf{r}^{\prime}).
\label{eq: CasidaK}
\end{equation}
The Coulomb integrals were computed using the LIBERI library \cite{Toyoda2010} which we modified to perform integrals using real spherical harmonics. Note that we have chosen to work within a linear-response framework because it allows exploitation of spherical symmetry which is broken by the perturbing electric field in a real-time framework.

Besides their energy, other important properties of neutral excitations are their oscillator strength $f_I$ and their collectivity $C_I$. The oscillator strength is obtained from the eigenvectors of Casida's equation via
\begin{equation}
f_{\rm I} = \frac{2m}{\hbar^2} \left| \sum_{vc} \sqrt{f_{vc}(\epsilon_{c}-\epsilon_{v}})\mu_{vc} F_{vc}^{I} \right| ^{2} \equiv \frac{2m}{\hbar}\omega_{I}\mu_{I}^{2},
\end{equation}
where $\mu_{vc}= e \int d\mathbf{r}z\phi_v(\mathbf{r})\phi_c(\mathbf{r})$ denotes the dipole moment matrix element for the vc-transition in the z-direction, which is parallel to the perturbing electric field, $\mu_{I}$ is the transition dipole moment of the excitation $I$ and $f_{vc}$ is the occupation number difference.

To define the collectivity of a neutral excitation, we first discuss the two extreme cases. When an excitation is perfectly collective, we expect that all components of $|F^I_{vc}|^2$ are equal to $1/N_{\rm pair}$, where $N_{\rm pair}$ denotes the total number of electron-hole pair states (note that this is finite as we only consider bound electron states). For an ideal electron-hole pair excitation, on the other hand, all components of $|F^I_{vc}|^2$ are zero except for one whose value is unity. For a general excitation, we first test if any component of $|F^I_{vc}|^2$ is larger than $\lambda/N_{\rm pair}$, where $\lambda$ is set to $500$ (we have tested that our results do not depend strongly on the choice of $\lambda$). If such a component is found, the state is identified as an electron-hole pair state and $C_{I}$ is set equal to the number of such components. Otherwise, the state is identified as collective and the collectivity is calculated as the total number of non-zero components of $|F^I_{vc}|^2$. In practice, we find that excitations with a plasmonic character have collectivity values corresponding to $\sim 10$ percent of non-zero components, while for electron-hole pairs typical $C_{I}$ is less than one percent. All DFT and TDDFT calculations were carried out using in-house computer codes.

\bibliography{library.bib}


\begin{addendum}
 \item The authors acknowledge support from the Thomas Young Centre under grant no. TYC-101. This work was supported through a studentship in the Centre for Doctoral Training on Theory and Simulation of Materials at Imperial College London funded by the EPSRC (EP/L015579/1) and  through EPSRC projects EP/L024926/1 and EP/L027151/1.

 \item[Competing Interests] The authors declare no competing interests.
 \item[Data availability] The data sets analyzed and the code used during the current study are available from the corresponding author on reasonable request.
 \item[Contributions]L. R. C. developed the computer code for these calculations.
L. R. C., O. H and J. L. contributed to the interpretation of the results and the writing of the manuscript.
 \item[Correspondence] Correspondence and requests for materials
should be addressed to L.R.C.~(email: lara.roman-castellanos14@imperial.ac.uk).
\end{addendum}

\end{document}


\maketitle
\section*{Supplementary note 1}

\textbf{The plasmon operator and its commutator}

The operators $b^\dagger_I$ in the effective Hamiltonian, see Eq. 1 of the main text, generate neutral excitations in state $I$ according to\cite{Yannouleas1992}
\begin{equation}
 \left| \Psi_{I} \right>   = b_{I}^{\dag} \left| GS \right>,   
\end{equation} 
where $\left| GS \right>$ denotes the correlated ground state of the many-electron system that satisfies $b_{I} \left| GS \right> =0$ for all $I$. In the random phase approximation, these operators can be expressed as linear combinations of electron-hole pairs
\begin{equation}
\label{eq: bosonicOperator}
 b_{I}^{\dag}   = \sum_{vc} X^{I}_{vc} c^{\dag}_{c}c_{v} - Y^{I}_{vc}c^{\dag}_{v}c_{c},
\end{equation} 
where $\mathbf{X}$ and  $\mathbf{Y}$ are the eigenvectors of the RPA pseudo-eigenvalue problem. For real orbitals, this is given by
\begin{equation}
\begin{pmatrix} 
A & B \\ 
-B & -A \\ 
\end{pmatrix}
\begin{pmatrix} 
X^{I} \\ 
Y^{I} \\ 
\end{pmatrix} = \hbar \omega_{I}\begin{pmatrix} 
X^{I} \\ 
Y^{I} \\ 
\end{pmatrix}.
\end{equation} 

In the above expression, we used $A_{vcv'c'} =\delta_{v'v}\delta_{c'c}(\epsilon_{c} - \epsilon_{v}) + K_{vcv'c'}$ and $B_{vcv'c'} = K_{vcv'c'}$, where $K_{vcv'c'}$ is the Coulomb integral defined in the main text. The pseudo-eigenvectors satisfy the following normalization condition\cite{Ullrich2014}
\begin{equation}
\label{eq: normalisation}
X^{I'}X^{I} - Y^{I'}Y^{I} =  \delta_{I'I}.
\end{equation}

Finally, we demonstrate that the $b_I$ operators fulfill the standard bosonic commutation relations \cite{Fetter1971}. Defining $z_{cv}^{\dag} \equiv c^{\dag}_{c}c_{v}$, we find


\begin{multline}
\left[ b_{I'}, b^{\dag}_{I} \right] = \sum_{v,c,v',c'} \left[ X^{I'}_{v'c'} z_{c'v'} - Y^{I'}_{v'c'} z^{\dag}_{c'v'} ,X^{I}_{vc} z^{\dag}_{cv} - Y^{I}_{vc} z_{cv} \right] = \\
\sum_{v,c,v',c'} -\left[ X^{I'}_{v'c'} z_{c'v'},Y^{I}_{vc} z_{cv} \right] -\left[ Y^{I'}_{v'c'} z^{\dag}_{c'v'},X^{I}_{vc} z^{\dag}_{cv} \right] + \left[ X^{I'}_{v'c'} z_{c'v'},X^{I}_{vc} z^{\dag}_{cv} \right]+\left[ Y^{I'}_{v'c'} z^{\dag}_{c'v'},Y^{I}_{vc} z_{cv} \right].
\end{multline} 
The first two terms are zero because of $[z_{vc},z_{v'c'}]=[z^\dagger_{vc},z^\dagger_{v'c'}]=0$ resulting in
\begin{equation}
\left[ b_{I'}, b^{\dag}_{I} \right] = \sum_{v,c,v',c'} \left[z_{c'v'},z^{\dag}_{cv} \right] \left(X_{v'c'}^{I'}X_{vc}^{I} - Y_{v'c'}^{I'}Y_{vc}^{I}  \right).
\end{equation}

Using the normalization condition, Eq.~\eqref{eq: normalisation}, and assuming that $\left< GS \left |z^{\dag}_{c'v'} z_{cv} \right|GS \right> \approx 0$, we find that


\begin{equation}
\left[ b_{I'}, b^{\dag}_{I} \right] = \delta_{I'I}.
\end{equation}

\section*{Supplementary note 2}

\textbf{Derivation of the expression for the electric field required to excite one plasmon}

In time-dependent linear response theory for a spin unpolarized system, the change in the electron density due to the external potential $\delta V(\mathbf{r},\omega)=-E_0z$ is given by 
\begin{equation*}
 \delta\rho(\mathbf{r},\omega) = \int d\mathbf{r}'\chi(\mathbf{r},\mathbf{r}',\omega) \delta V(\mathbf{r^{\prime}},\omega),   
\end{equation*}
where $\chi(\mathbf{r},\mathbf{r}',\omega)$ denotes the interacting susceptibility. The (retarded) susceptibility can be written as a sum over normal modes
\begin{equation*}
    \chi(\mathbf{r},\mathbf{r}',\omega) = \sum_I \rho_I(\mathbf{r})\rho_I(\mathbf{r}')\left[ \frac{1}{\hbar (\omega-\omega_I) + i\gamma_I} - \frac{1}{\hbar (\omega+ \omega_I)+i\gamma_I}\right],
\end{equation*}
where $\rho_I(\mathbf{r})$, $\omega_I$ and $\gamma_I$ denote the transition density, the transition energy and the linewidth of the $I$-th neutral excitation, respectively. 

Close to the plasmon energy $\omega \approx \omega_P$, the contribution from the plasmon mode dominates the susceptibility\cite{Bursi2016}, i.e. $\chi(\mathbf{r},\mathbf{r}',\omega) \approx  \rho_{P}(\mathbf{r})\rho_{P}(\mathbf{r}')\frac{1}{i\gamma_{P}}$
and the induced density at the plasmon energy is given by 
\begin{equation*}
 \delta\rho(\mathbf{r},\omega_P) =-\frac{E_0}{\gamma_P} \rho_P(\mathbf{r}) \int d\mathbf{r}'z'\rho_P(\mathbf{r}') \equiv -\frac{E_0}{\gamma_p}\mu_P \rho_P(\mathbf{r}),
\end{equation*}
where $\mu_P$ denotes the plasmon transition dipole matrix element. 
According to Eq.~(4), $\rho_P(\mathbf{r})$ is the transition density of a \emph{single plasmon excitation}. Therefore, we find that a field strength 
\begin{equation*}
    E_0^{1PL} = \frac{\gamma_P}{\mu_P}
\end{equation*}
results in the excitation of a single plasmon.

\section*{Supplementary note 3}

\textbf{Additional insights into the semiclassical model}

\begin{figure}[htpb]
\centering
\includegraphics[width=300pt]{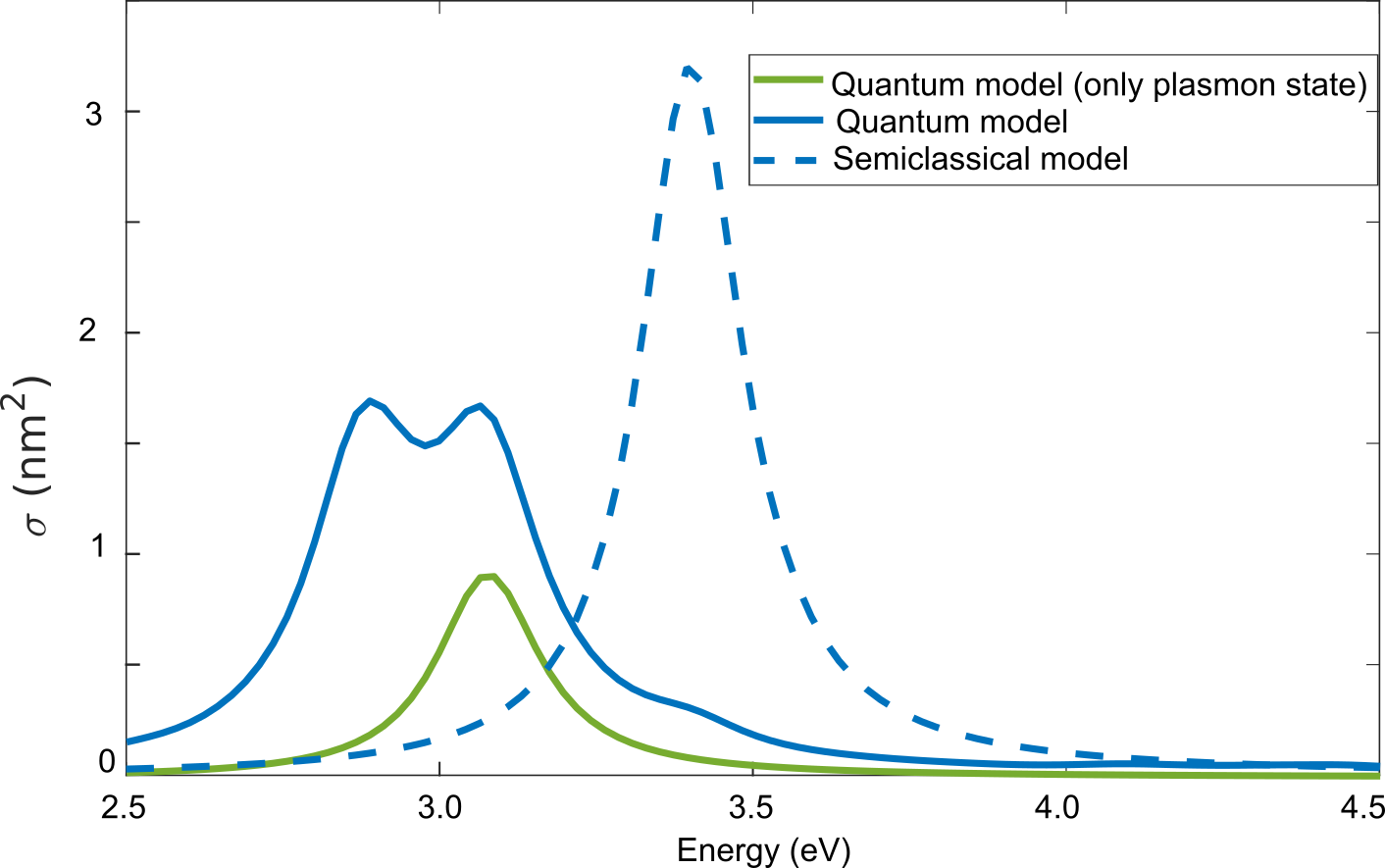}
\caption{Optical absorption spectra calculated using the quasistatic approximation (blue dashed line) and quantum-mechanical TDDFT (blue continuous line) for $Na_{92}$. The green line shows the TDDFT spectrum that is obtained when only absorption by the plasmonic state is allowed.}
\label{fig: ABS}
\end{figure}

Supplementary Figure \ref{fig: ABS} compares the absorption spectrum of the Na$_{92}$ nanoparticle from the quasistatic approximation \cite{Maier2001} (using the bulk dielectric function described in the main text) with the quantum-mechanical TDDFT result. We also show the TDDFT absorption spectrum that is obtained when only the plasmonic states are taken into account. The difference with the full TDDFT result demonstrates that a significant amount of oscillator strength is contained in non-plasmonic excited states. In contrast, all the oscillator strength which is available according to the f-sum rule is contained in the plasmon peak in the classical result.

\begin{figure}[h!]
\centering
\includegraphics[width=300pt]{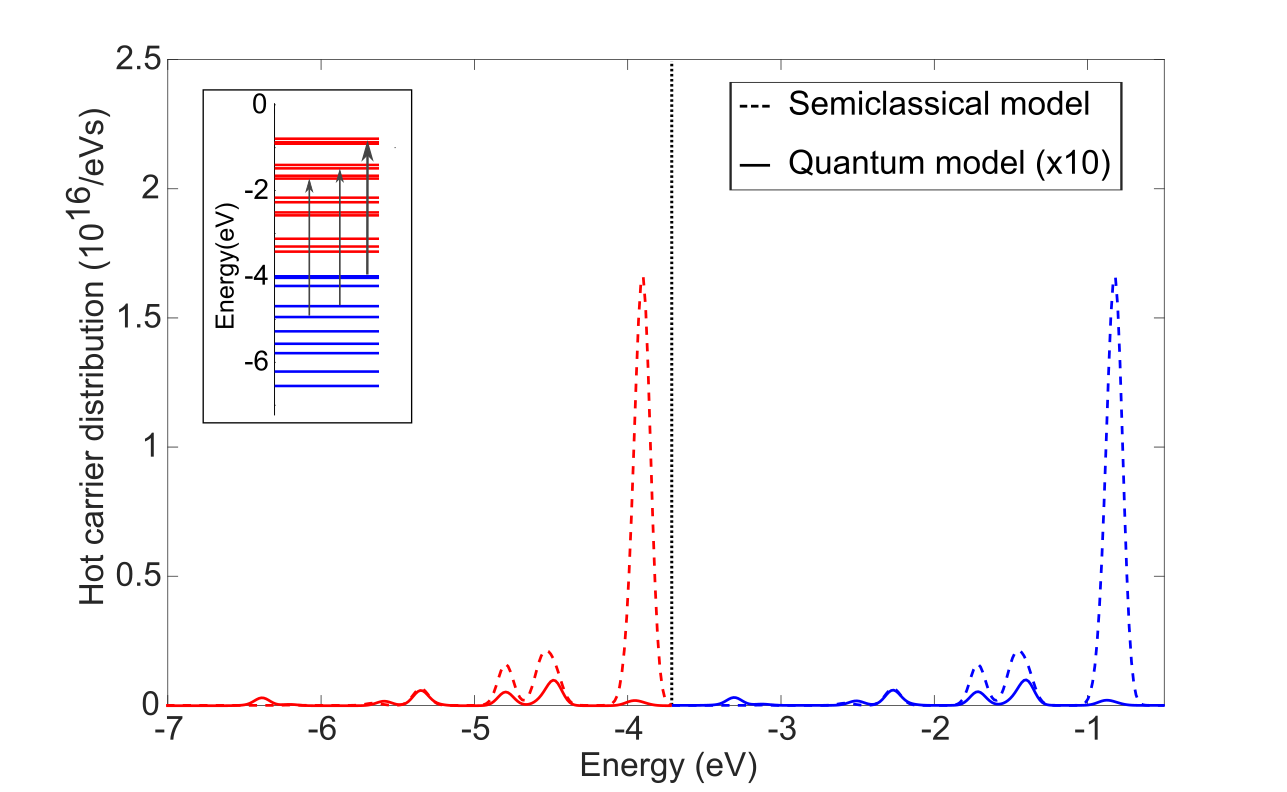}
\caption{Comparison of the energy distributions of hot electrons (blue lines) and hot holes (red lines) in Na$_{92}$ obtained from the semiclassical approach (dashed lines) and the new quantum-mechanical method (solid lines). Note that the latter has been scaled by a factor of 10. The inset shows the relevant electronic transitions from occupied (blue) to empty (red) levels for the semiclassical model.}
\label{fig: HCDSC}
\end{figure}

The larger plasmon oscillator strength in the semiclassical approach gives rise to an enhanced hot-electron generation. Supplementary Figure~\ref{fig: HCDSC} compares the distribution of hot electrons and holes in Na$_{92}$ from the semiclassical approach to the result of the new quantum-mechanical approach (see main text for a detailed description of these approaches). Note that the semiclassical approach differs from the quantum-mechanical theory only in the value of the electron-plasmon coupling strength. Therefore, the peaks in the distribution functions are located at the same energies in both approaches, but the height of the peaks is significantly larger in the semiclassical case. Despite its quantitative inaccuracy, it is interesting to note that the semiclassical approach can yield qualitative insights. For example, it predicts correctly that a larger amount of energy is deposited into hot electrons than into hot holes.

\begin{figure}[h!]
\centering
\includegraphics[width=400pt]{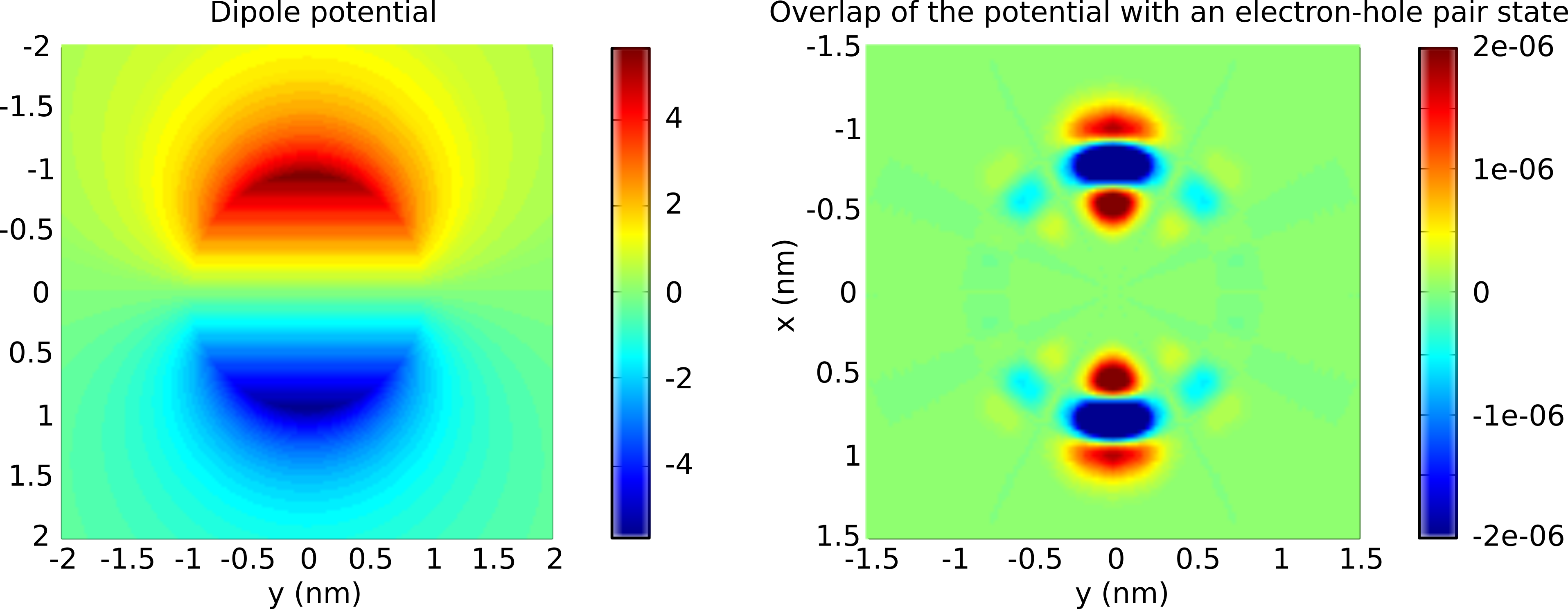}
\caption{Left: Plasmon potential (in volts) from the semiclassical approach (calculated using Eq.10 of the main manuscript). Right: Overlap of the semiclassical dipole potential with an electron-hole pair $\phi^*_{c}(\mathbf{r})\phi_{v}(\mathbf{r})$.}
\label{fig: SCPot}
\end{figure}

\begin{figure}[h!]
\centering
\includegraphics[width=400pt]{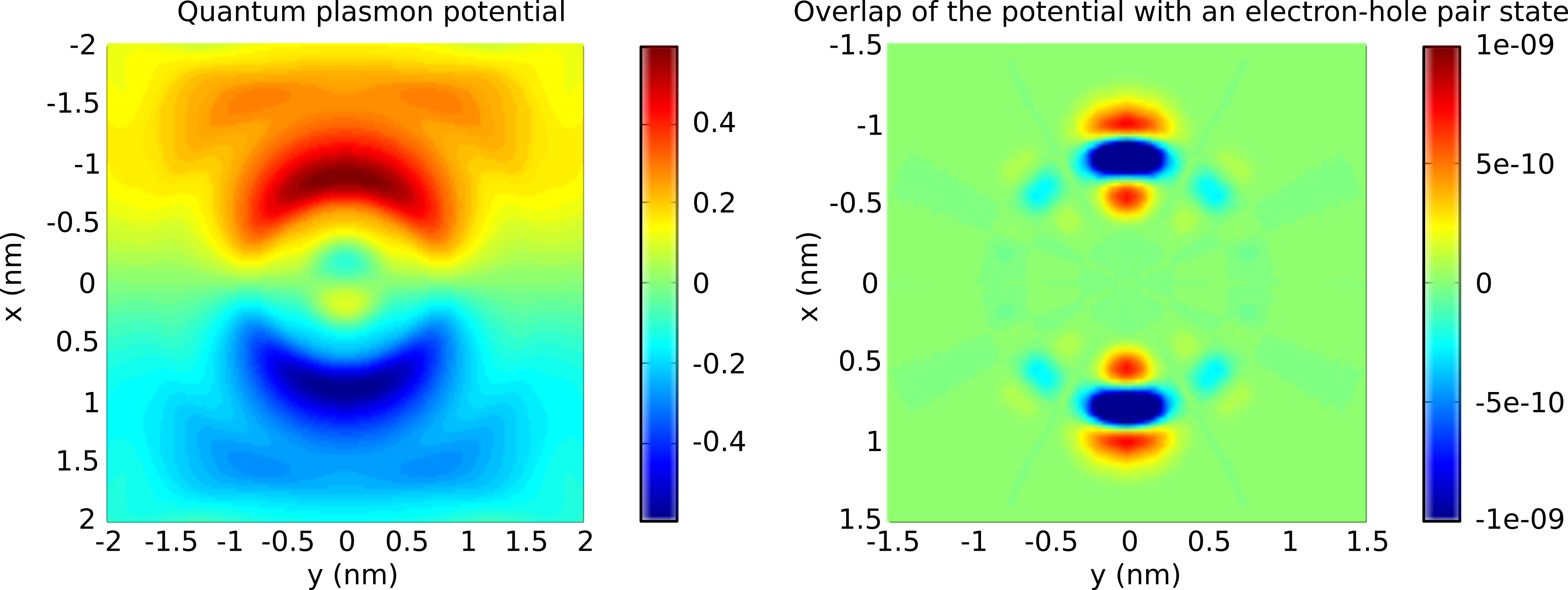}
\caption{Left: Plasmon potential (in volts) calculated using the new quantum-mechanical approach. Right: Overlap of the quantum-mechanical plasmon potential with an electron-hole pair $\phi^*_{c}(\mathbf{r})\phi_{v}(\mathbf{r})$.}
\label{fig: QuantumPot}
\end{figure}

To understand the discrepancy between the hot-carrier generation rates from the quantum-mechanical and the semiclassical calculations, we have calculated the plasmon potentials for the two approaches. Supplementary Figures \ref{fig: SCPot} and \ref{fig: QuantumPot} (left panels) show the 
plasmon potential for a Na$_{92}$ nanoparticle from the semiclassical approach and quantum-mechanical theory. While the semiclassical result has a purely dipolar shape, the quantum-mechanical result has additional structure resulting from the more complicated plasmon charge distribution, see discussion in the main text. Morever, the semiclassical potential is significantly stronger than the quantum potential. We have also calculated the product of the plasmon potential and the electron-hole pair state $\psi^*_c(\mathbf{r})\psi_v(\mathbf{r})$ for the transition that gives rise to the main peak of the semiclassical hot-carrier distribution in Supplementary Figure \ref{fig: HCDSC}. Supplementary Figures \ref{fig: SCPot} and \ref{fig: QuantumPot} show that the overlap is significantly larger for the semiclassical potential explaining larger electron-plasmon coupling strengths and increased hot-carrier generation rates.

\newpage

\medskip

\bibliography{library.bib}